\documentclass[amssymb,amsmath,english,aps,showpacs,twocolumn,floatfix]{revtex4}
\usepackage{hyperref}
\usepackage{subfigure}
\usepackage{graphicx}

\newcommand{\unit}[1]{\,{\rm #1}}
\newcommand{\Rm}{Re_{\rm m}}
\providecommand{\boldsymbol}[1]{\mbox{\boldmath $#1$}}

\providecommand{\tabularnewline}{\\}

\usepackage{babel}

\begin{document}

\preprint{APS/123-DPP}

\title{Traveling waves in magnetized Taylor-Couette flow}

\author{Wei Liu$^{1}$\footnote{Email: wliu@pppl.gov}, Jeremy Goodman$^{2}$, Hantao Ji$^{1}$}

\affiliation{$^{1}$Center for Magnetic Self-Organization in Laboratory and Astrophysical Plasmas, Princeton Plasma Physics Laboratory, Princeton University, P.O. Box
451, Princeton, NJ 08543 }

\affiliation{$^{2}$Princeton University Observatory, Princeton, NJ 08544}

\date{\today{}}

\begin{abstract}

 We investigate numerically a traveling wave pattern observed in experimental  
 magnetized Taylor-Couette flow
 at low magnetic Reynolds number. By accurately modeling viscous and  
 magnetic boundaries in
 all directions, we reproduce the experimentally measured wave patterns and their  
 amplitudes.  Contrary to previous claims,  
 the waves are shown to be transiently amplified disturbances launched by 
 viscous boundary layers
 rather than globally unstable magnetorotational modes.
  

\end{abstract}

\maketitle

\section{Introduction}\label{sec:intro}
The luminosity of most astrophysical accretion disks 
probably depends upon the magnetorotational instability (MRI)\cite{bh98},
which has inspired searches for MRI in Taylor-Couette flow. 
Standard MRI modes will not grow unless both the rotation period and
the Alfv\'en crossing time are shorter than the timescale for magnetic
diffusion.  This requires that both the magnetic Reynolds number
$\Rm\equiv\Omega_{1}r_{1}(r_{2}-r_{1})/\eta$ and the Lundquist
number $S\equiv V_{Az}^{0}(r_{2}-r_{1})/\eta$ be $\gtrsim 1$, where
$V_{Az}^{0}=B_{z}^{0}/\sqrt{4\pi\rho}$ is the Alfv\'en speed and
$B_{z}^{0}$ is the background magnetic field parallel to the
angular velocity $\boldsymbol{\Omega}=\Omega\boldsymbol{e}_z$.
No laboratory study of standard MRI has been completed except for
that of \citet{sisan04}, whose experiment proceeded from a background
state that was already hydrodynamically turbulent before the field was applied.
Recent linear analyses of axially periodic or infinite magnetized
Taylor-Couette flow
has shown that MRI may grow at much reduced $\Rm$ and $S$ in the
presence of a combination of axial and current-free toroidal field
\cite{hr05,rhss05}.
We call such modes ``helical'' MRI (HMRI)

The PROMISE (Potsdam Rossendorf Magnetic Instability Experiment) group have claimed to observe HMRI experimentally
\cite{sgg06, rhsgg06, sgg07}. At magnetic and flow parameters where
linear analysis predicts instability, persistent fluctuations were
measured that appeared to form axially traveling waves, consistent
with expectations for HMRI.  Similar behavior has been seen in
nonlinear numerical simulations that approximate the experimental
conditions, including realistic viscous boundary conditions for the
velocities, but simplified ones for the magnetic field: perfectly
conducting cylinders, and pseudo-vacuum conditions at the endcaps,
when present \cite{sgg07,sr06}.  Both axially periodic
and finite cylinders showed unsteady flow, the former case being more
regular. However, the nonlinear simulations in \cite{sgg07,sr06} used 
somewhat different values for the cylinder rotation rates and other 
parameters than those reported in \cite{sgg06}.

Previously, however, we have raised doubts about both the experimental
realizability of HMRI and its astrophysical relevance\cite{lghj06}.
Finite cylinders with insulating endcaps were shown to reduce the
growth rate and to stabilize highly resistive flows entirely, at least
inviscid ones.

Here we report nonlinear simulations with the ZEUS-MP
2.0 code \cite{hnf06}, which is a 
time-explicit, compressible, astrophysical ideal MHD parallel 3D code,
to which we have added viscosity, resistivity (with subcycling to reduce
the cost of the induction equation), and vacuum boundary
conditions, for axisymmetric flows in cylindrical coordinates
$(r,\varphi,z)$\cite{lgj06}.
The parameters of PROMISE as reported in or inferred from
\cite{sgg06} are used: gallium density
$\rho=6.35\unit{g\;cm^{-3}}$, magnetic diffusivity $\eta=2.43\times10^{3}\unit{cm^{2}\;s^{-1}}$, magnetic Prandtl
number $Pr_{\rm m}\equiv\nu/\eta=1.40\times10^{-6}$; Reynolds number $Re\equiv
\Omega_{1}r_{1}(r_{2}-r_{1})/\nu=1775$; axial current
$I_{z}=6000\unit{A}$; toroidal-coil currents $I_\varphi=0, 50, 75,
120\unit{A}$; and dimensions as in Fig.~\ref{boundary}. 
For the first time,
the finite conductivity and thickness of the
copper vessel are 
allowed for ($\eta_{\rm Cu}= 1.335\times10^{2}\unit{cm^{2}s^{-1}}$),
and this noticeably improves agreement with the measurements compared
to previous linear calculations with radially perfectly conducting, axially 
periodic boundaries \cite{sgg06, rhsgg06}.
Please note the difference of the direction of $\Omega$, $B_{z}$ and $B_{\varphi}$ (components 
measured in a right handed coordinate system) between
this paper, where they are all assumed to be positive, and the experimental setup 
presented in \cite{sgg06}, where they are all negative (private communication). 
The direction of the traveling wave depends on the sign of the Poynting flux defined
as $-r\Omega B_{\varphi}B_{z}/\mu_{0}$ \cite{lghj06}. Thus the direction of the traveling
wave reported here is opposite as reported in \cite{sgg06}.

\section{Boundary Conditions}\label{sec:bndry} 
\begin{figure}[!htp]

\caption{
  Computational domain for simulations of PROMISE experiment. Region
  (I): Inner copper cylinder, angular velocity $\Omega_{1}$.
  (II): outer copper cylinder and bottom endcap, 
  $\Omega_{2}$. (III): liquid gallium; (IV):
  vacuum. Thick dashed line: insulating upper endcap, $\Omega=0$. Dimensions:
  $r_{1}=4.0\unit{cm}$; $r_{2}=8.0\unit{cm}$; $h=40.0\unit{cm}$;
  $d_{wI}=1.0\unit{cm}$; $d_{wII}=1.5\unit{cm}$;
  $\Omega_{1}/2\pi=3.6\unit{rpm}$; $\Omega_{2}/2\pi=0.972\unit{rpm}$.
  The exact configuration of the toroidal coils being unavailable to us,
  six coils (black rectangles) with dimensions as shown were used,
  with $67$ turns in the two coils nearest the midplane and $72$ in the rest.
  Currents $I_\varphi$ were adjusted to reproduce the reported \cite{sgg06}
  Hartmann numbers $Ha\equiv B_{z}^{0}r_{1}/\sqrt{\rho\mu_{0}\eta
  \nu}$.
\label{boundary} }

\scalebox{0.4}{\includegraphics{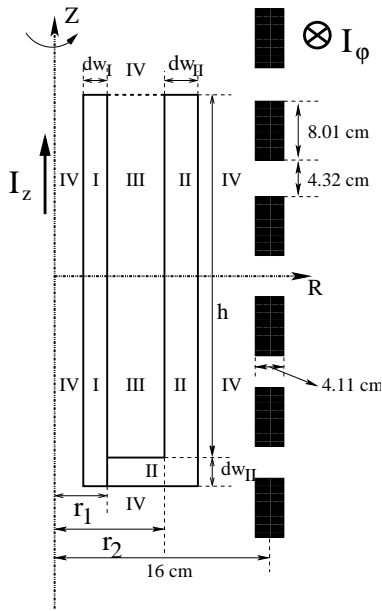}}
    
\end{figure}

At the low frequencies relevant to PROMISE ($f\sim0.01\unit{Hz}$),
the skin depth of Copper $\delta_{w}=\sqrt{\eta_{\rm Cu}/\pi
  f\mu_{0}}\approx19\unit{cm}$, which is much larger than the
thickness of the copper vessel surrounding the gallium in the PROMISE
experiment, $d_{w}\approx1.0\unit{cm}$, so that the magnetic field
diffuses rather easily into the boundary.  On the other hand, if one
considers axial currents, the gallium and the copper wall act as
resistors in parallel; taking into account their conductivities and
radial thickness, one finds that their resistances are comparable
[$R_{I}:R_{II}:R_{III}=3:1:9$; see Fig.~\ref{boundary} for the
subscripts].  Therefore, the currents carried by the copper walls
could be important for the toroidal field, and a perfectly insulating
boundary condition is also inappropriate.

We have adapted a linear axisymmetric
code developed by \cite{gj02,lghj06} to allow
for a helical field.  Vertical periodicity is assumed, to allow
separation of variables, but the full viscous and resistive radial
equations are solved using finite differences, and a variety
of radial boundary conditions can be imposed.
For perfectly conducting boundaries
and $I_\varphi=75\unit{A}$, 
where \cite{sgg06} report persistent waves, our code indeed finds
a complex growth rate: $s\approx0.0057+0.057i\unit{s^{-1}}$.
But for insulating boundaries,
the same parameters yield stability.

This analysis points to the need for boundary conditions that
accurately represent the influence of the copper vessels on the field.
In the linear code just mentioned, we use the thin-wall approximation
of \cite{mb00}, which in effect treats the cylinders as insulating for
the poloidal field but conducting for the toroidal field.  The errors
of this approximation increase with the ratio  of
wall thickness to gap width, which is not very small
($\approx 0.25$) in our case.
Growth is predicted, but at a smaller rate than for
perfectly conducting walls, $s\approx0.0052+0.056\,i~\unit{s^{-1}}$.
The insensitivity of the imaginary part to the magnetic boundaries
supports the interpretation that these
modes are hydrodynamic inertial oscillations weakly destabilized by
the helical field \cite{lghj06}.

In our nonlinear simulations, we include the copper walls (regions I
and II) in the computational domain
(Fig.~\ref{boundary}), but not the external coils themselves, whose
inductive effects are therefore neglected.  Outside the walls
(region IV) we match onto a vacuum field
$\boldsymbol{B}_{\rm ext}=\boldsymbol{\nabla} \Phi$ vanishing at
infinity.  This is relatively straightforward in spherical geometry
(used by many geodynamo experiments) because Laplace's equation
separates. Our case is more difficult, because while Laplace's
equation separates in cylindrical coordinates when the boundary is an
infinite cylinder, it does not fully separate outside a \emph{finite}
cylinder. Therefore we use an integral formulation that does
not assume separability.
The idea, called von~Hagenow's method \cite{lk76},
is to find a surface current on the boundary that is equivalent
to the current density in the interior as the source
for $\boldsymbol{B}_{\rm ext}$ \emph{via}
the free-space Green's function.
The surface current is obtained by first solving
the Grad-Shafranov equation \cite{gr58,sv66} in the interior
with \emph{conducting} boundary conditions, 
a problem that \emph{is} separable in our case and is solved efficiently
by combining FFTs along $z$ with tridiagonal matrix inversion along $r$.

\section{Results and Discussion}\label{sec:resu}

We start with purely hydrodynamic (unmagnetized) simulations. For
$\mu\equiv\Omega_{2}/\Omega_{1}=0.27$, what we see is simply an Ekman
flow driven by the top and bottom end plates.  Due to the stronger
pumping at the upper, stationary lid, the two Ekman cells are of unequal
size. They are separated vertically by a narrow, intense radial
outflow, hereafter the ``jet'', lying at about $11\unit{cm}$ above
the bottom endcap. As discussed in \cite{kjg04}, the jet is
unsteady at $Re\gtrsim 10^3$; it flaps or wanders rapidly in
the poloidal plane. This has been verified by the PROMISE
group (private communication). The amplitude of
the flapping is $\pm0.4\unit{mm\;s^{-1}}$.



    

\begin{figure}[!htp]

\caption{
  (color). Axial velocities $[\unit{mm\;s^{-1}}]$ versus time and
  depth sampled $1.5\unit{cm}$ from the outer cylinder, for
  the parameters of the PROMISE experiment \cite{sgg06} 
  with toroidal currents $I_\varphi$ as marked. Note height increases upward from the bottom endcap. 
  No-slip velocity boundary conditions are imposed at the rigidly rotating
  endcaps, but the steady part of the resulting Ekman circulation is
  suppressed in these plots by subtracting the time average at each height.
  The waves appear to be
  absorbed near the Ekman jet, at $z\approx 100\unit{mm}$.
  \label{wave} }

\subfigure{\scalebox{0.4}{\includegraphics{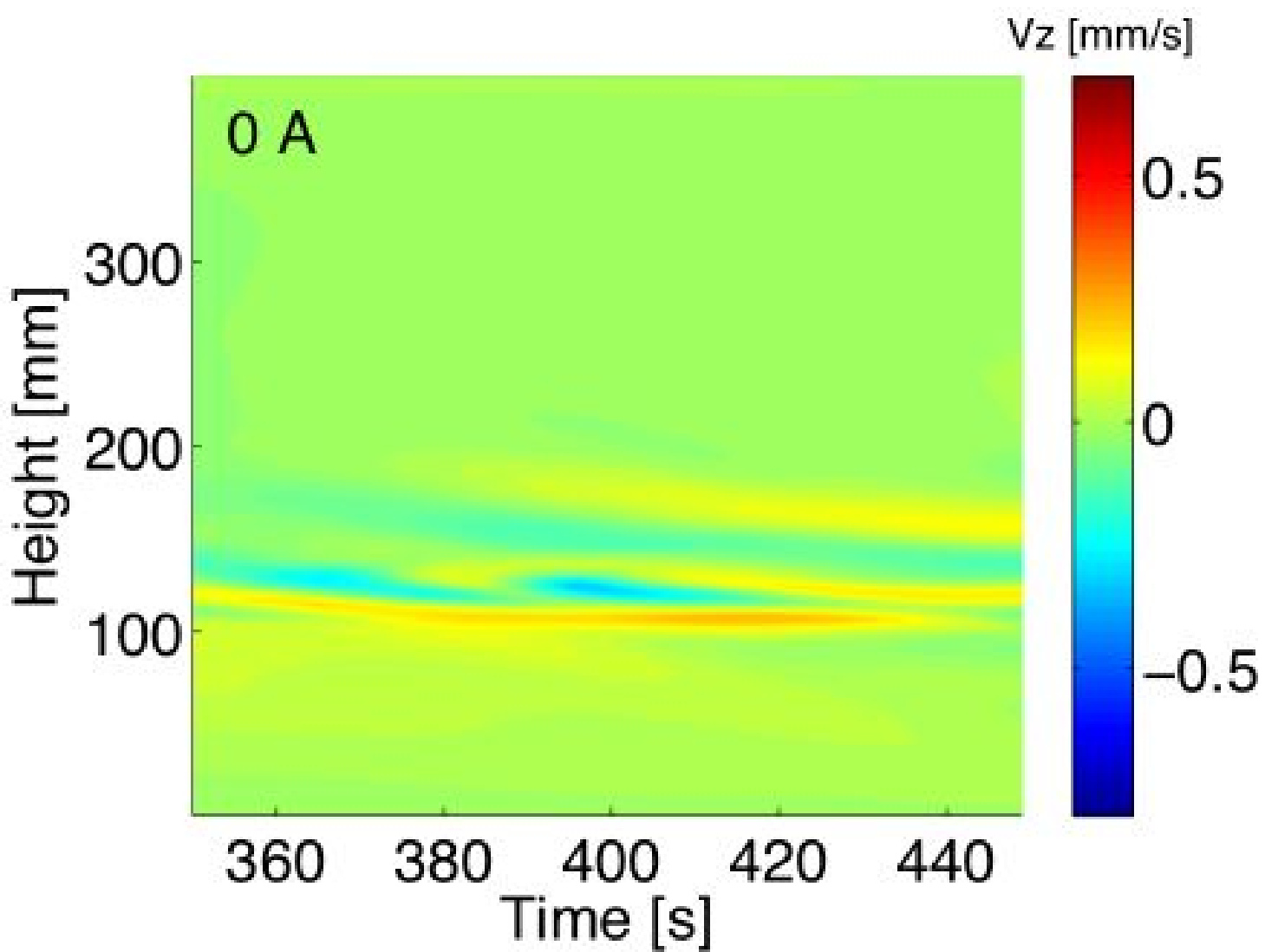}}}$\;$
\subfigure{\scalebox{0.4}{\includegraphics{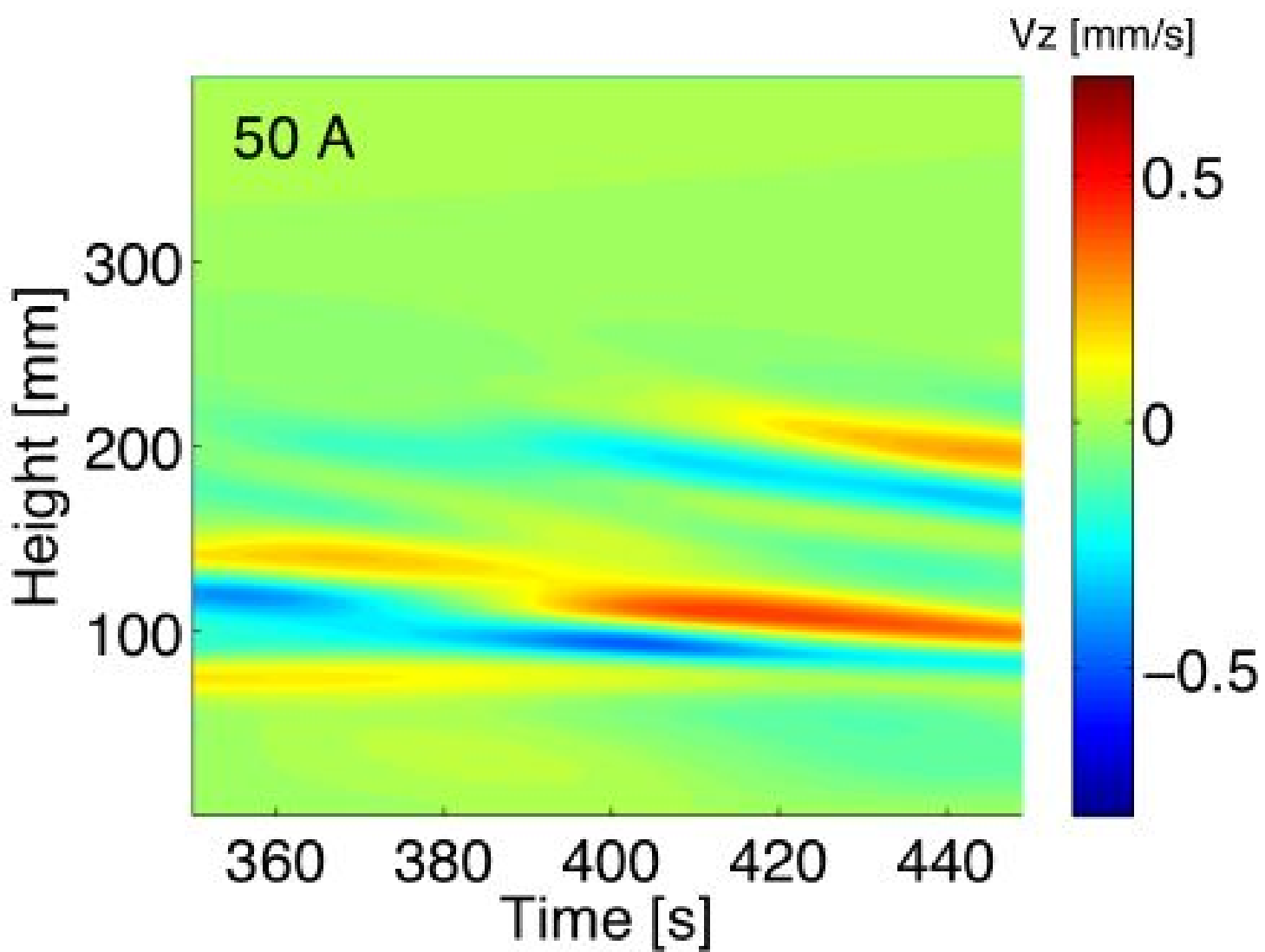}}}$\;$
\subfigure{\scalebox{0.4}{\includegraphics{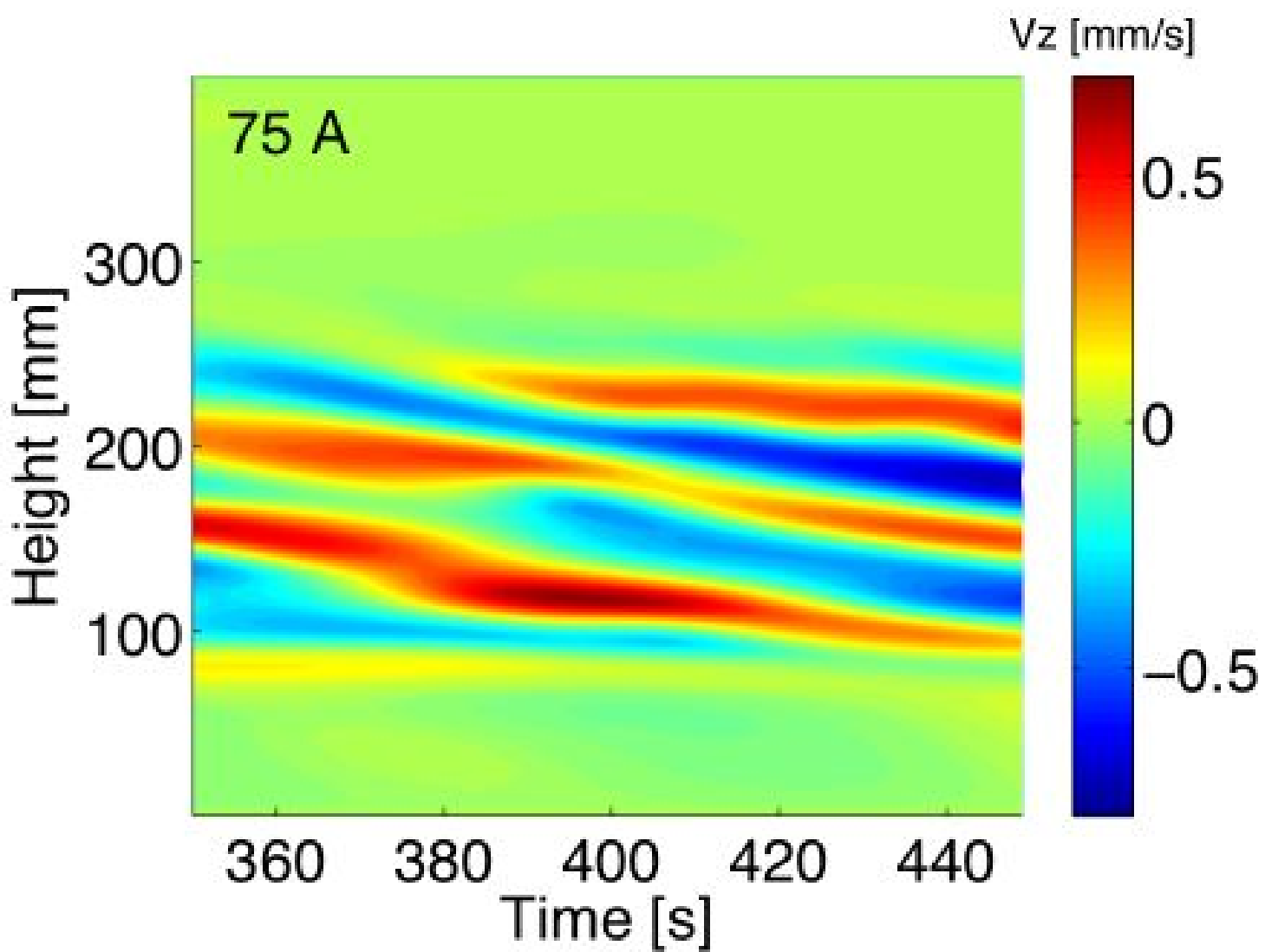}}}$\;$
\subfigure{\scalebox{0.4}{\includegraphics{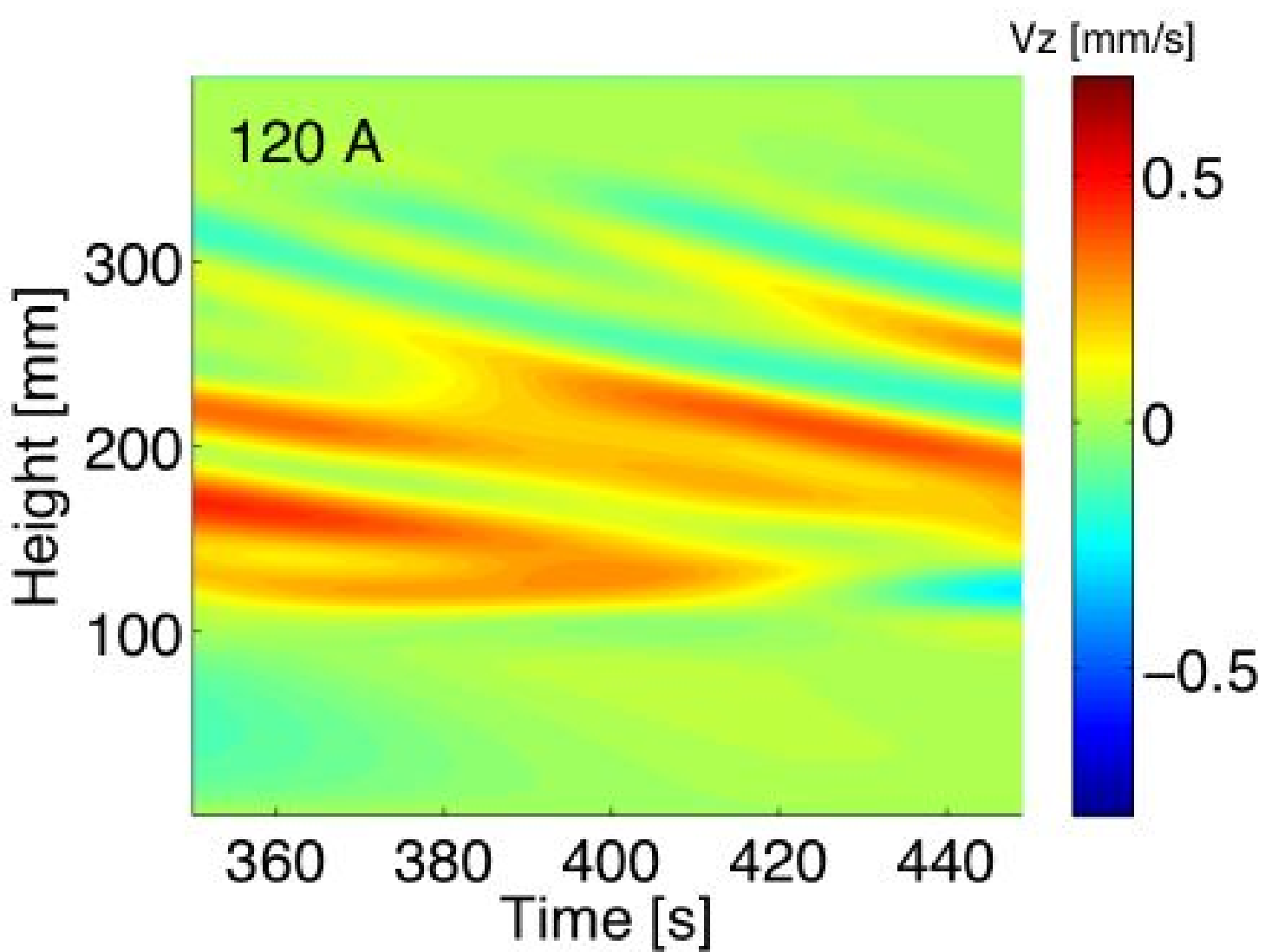}}}
    
\end{figure}

\begin{figure}[!htp]

\caption{
  (color). Panel~(a): An extended version of the case
  $I_\varphi=75\unit{A}$ shown in Fig.~\ref{wave} but without
  subtraction of the time average.  The two Ekman cells are visible
  as the upflow (orange) at $z>100\unit{mm}$ and downflow (blue)
  at $z<100\unit{mm}$; these are the expected directions of flow near
  the outer cylinder.
  Panel~(b): The same case again, except that after $t=360\unit{s}$, 
  the no-slip boundary condition at both endcaps is changed to an
  ideal Couette profile, \emph{i.e.} 
  $\Omega(r)=a+br^{-2}$ with $a$ and $b$ chosen to make $\Omega$
  continuous at both cylinders; this eliminates Ekman circulation.
  Thereafter, the wave seems to be absorbed near the bottom
  ($z\approx0\unit{mm}$) rather than the jet $(z\approx100\unit{mm})$, which
  itself dies out after $t\approx395\unit{s}$.
  \label{decay} }

\subfigure{\scalebox{0.4}{\includegraphics{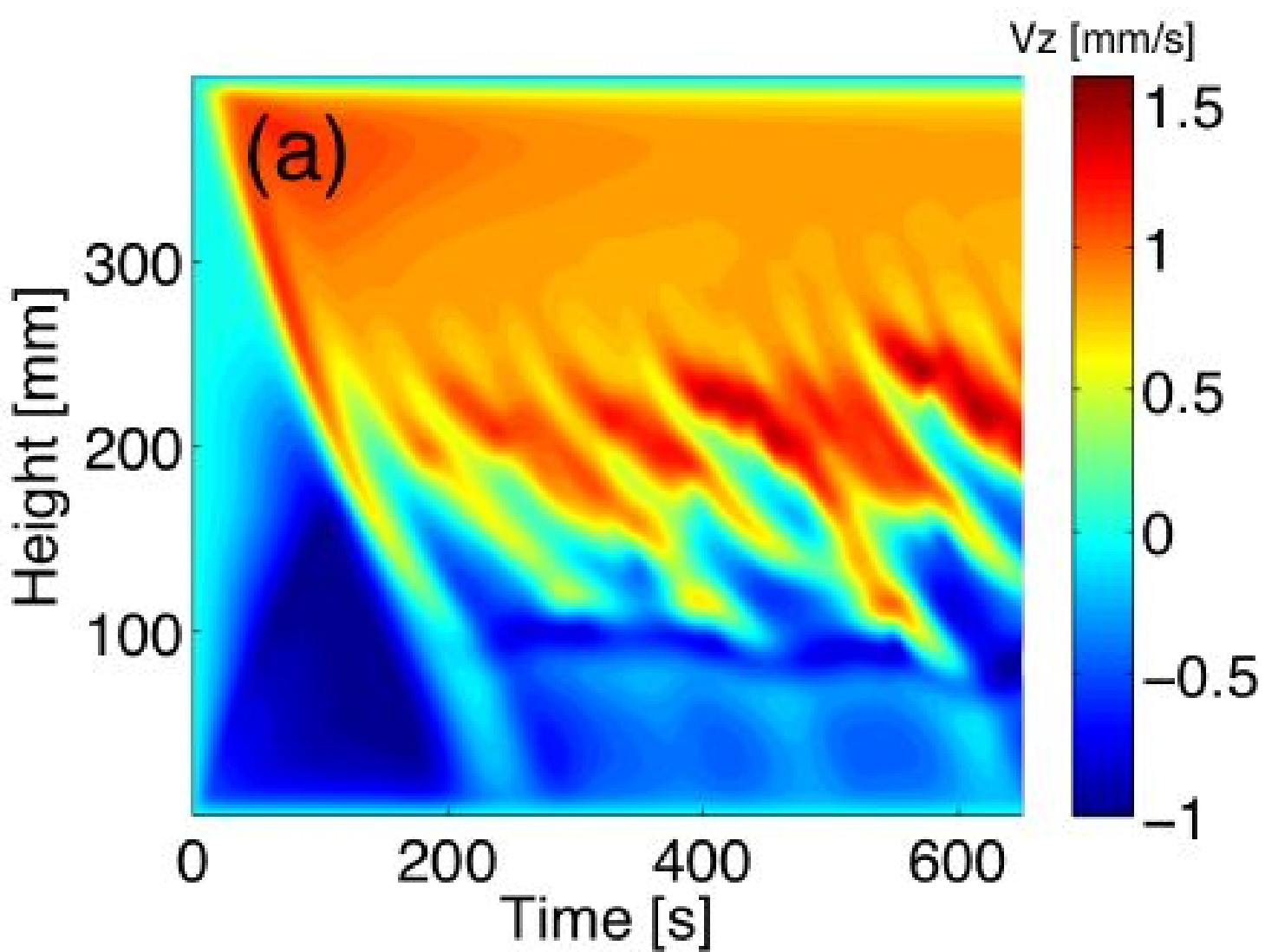}}}$\;$
\subfigure{\scalebox{0.4}{\includegraphics{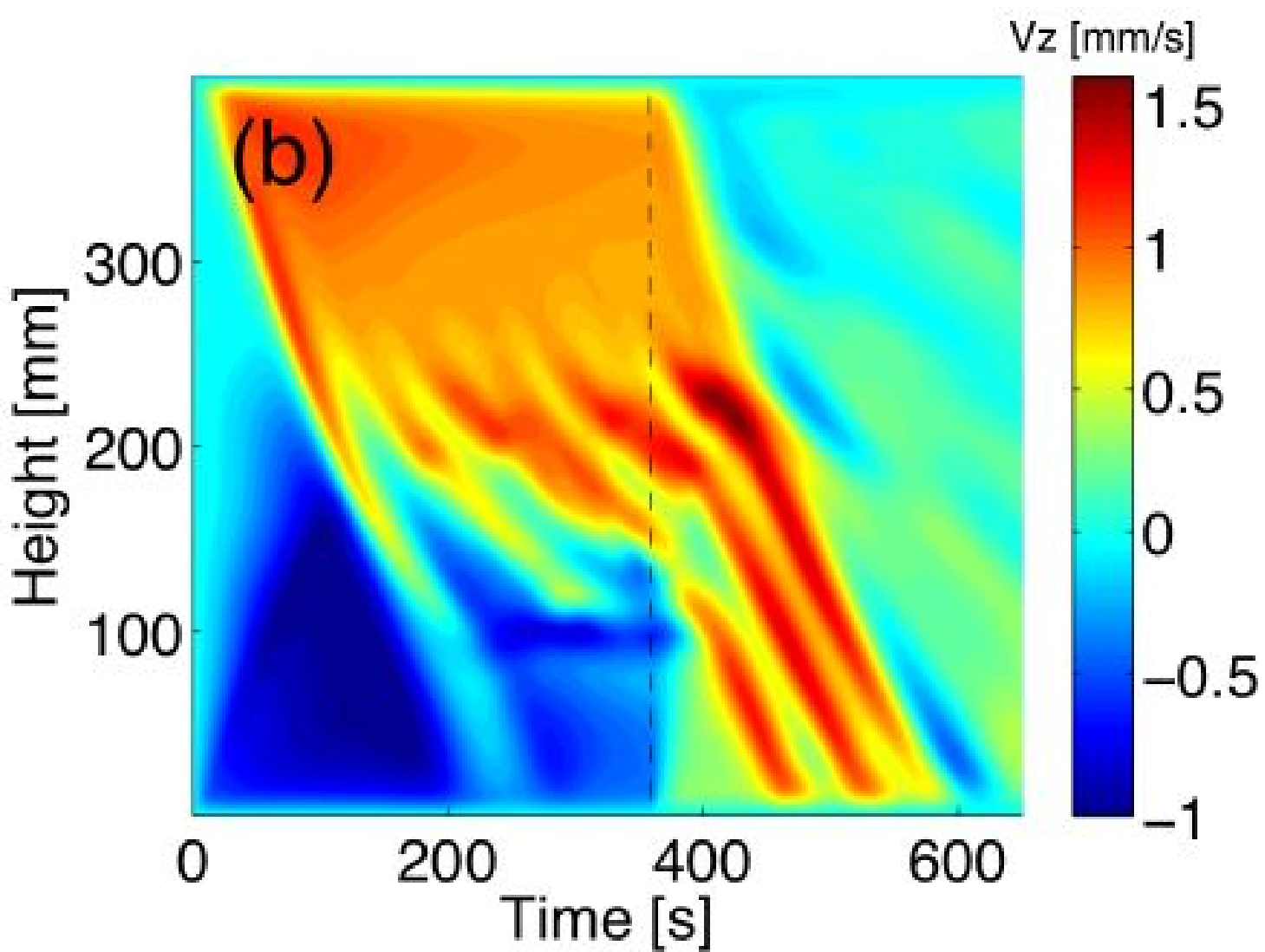}}}
    
\end{figure}

Background states with purely axial or
purely azimuthal magnetic fields are symmetric under reflection
$z\to-z$, but a helical field breaks this
symmetry\cite{ke96}. As a result, growing modes in
vertically infinite or periodic cylinders propagate axially
in a unique direction: that of the
background Poynting flux $-r\Omega B_\theta B_z/\mu_0$
\cite{lghj06}. Fig.~\ref{wave} displays vertical velocities
near the outer cylinder in simulations corresponding
to the experimental runs of \cite{sgg06} for several values
of the toroidal current, $I_\varphi$.
A wave pattern very similar to that in the
experimental data is seen.  It is most obvious for
$I_\varphi=75\unit{A}$, just as in the experiment.
Considering that we do not use exactly the same external coil
configuration as PROMISE, the agreement is remarkably good
(Table.~\ref{table1}). 
 \begin{table}[!htp]

\caption{\label{table1} Comparison of results for
the frequency, wavelength, axial phase speed, and amplitude
obtained from simulation and experiment 
for the case $I_\varphi=75\unit{A}$.  $f_1\equiv\Omega_1/2\pi$ is rotation frequency
of inner cylinder.}
\begin{tabular}{|c|c|c|c|c|}
\hline 
&
Calculation of  \cite{sgg06,rhsgg06}&
Experiment &
Our Simulation  \tabularnewline
\hline
\hline 
$f_{\rm wave}/f_{1}$&
$\sim0.14$&
$\sim0.15$&
0.15\tabularnewline
\hline 
$\lambda_{wave}$~[cm] &
$\sim12$&
6&
6\tabularnewline
\hline 
$v_{p}$[$\unit{mm\;s^{-1}}$]&
1.1&
0.8&
0.7\tabularnewline
\hline 
$A$[$\unit{mm\;s^{-1}}$]&
unavailable&
$\gtrsim 0.4$&
$\gtrsim 0.6$\tabularnewline
\hline
\end{tabular}
\end{table}

Interestingly, the jet becomes
nearly steady when $I_\varphi\ge 50\unit{A}$.
It is known that
Ekman circulation is significantly modified when the
Elsasser number $\Lambda\equiv
B^{2}/(\mu_{0}\rho\eta\Omega)\gtrsim 1$\cite{gp71}.
If we use $|\boldsymbol{B}(r_1)|$ for the field strength
and $\Omega_2$ for $\Omega$ in this expression, then
$\Lambda=4.8$ at $I_\varphi=75\unit{A}$.  

On the other hand,
the magnetic
field clearly promotes unsteadiness in the interior flow.
The waves seen in Fig.~\ref{wave} are probably related
to HMRI, but we do not believe that they arise from a global
instability of the experimental Couette flow.  To demonstrate this,
we have repeated the third ($I_{\varphi}=75\unit{A}$) simulation
shown in Fig.~\ref{wave} with different velocity boundary conditions.
First, when we replace the rigidly rotating endcaps with differentially
rotating ones that follow the ideal angular velocity profile
of an infinitely long Taylor-Couette flow, then instead of the
persistent travelling waves seen in Fig.~\ref{wave}, we see slowly
damping standing waves, which we interpret as
inertial oscillations excited by a small numerical force imbalance
in the inital conditions\cite{lghj06}.  Second, we perform
a simulation that begins with the experimental boundary conditions until
the traveling waves are well established, and then switches abruptly
to ideal-Couette endcaps.  After the switch, the Ekman circulation
stops and the traveling waves disappear after one axial propagation
time, as if they had been emitted by the Ekman layer at the upper endcap
or by the layers on the upper part of the cylinders (Fig.~\ref{decay}).
After the switch in boundary conditions but before the waves fully
disappear, their vertical phase speed increases
from $-0.7\unit{cm\;s^{-1}}$ to
$-1.1\unit{cm\;s^{-1}}$; the latter is the speed predicted by 
linear analysis for axially periodic flow \cite{rhsgg06}
(Fig.~\ref{decay}). Both numerical tests support the
interpretation that the
  wave pattern observed in the simulation and in the experiment is not a
  global HMRI mode but rather a transient disturbance that is somehow
excited by the Ekman circulation and then transiently amplified
as it propagates along the background axial Poynting flux, but is
then absorbed once it reaches the jet or the bottom end cap.


The authors would like to thank James Stone for the advice on the ZEUS
code and thank Stephen Jardin for the advice of implementing full insulating boundary condition. This work was supported by the US Department of Energy, NASA
under grants ATP03-0084-0106 and APRA04-0000-0152 and also by the
National Science Foundation under grant AST-0205903.



\end{document}